\documentclass[a4paper, 10pt,onecolumn]{ieeeconf}      

\IEEEoverridecommandlockouts                              
 
\usepackage{amsmath}
\usepackage{amssymb}
\usepackage{graphicx}
\usepackage[utf8]{inputenc}
\usepackage[dvipsnames]{xcolor}
\usepackage{tikz}
\usepackage{siunitx}
\usepackage[hidelinks]{hyperref}

\usepackage{subcaption}
\DeclareCaptionLabelSeparator{periodspace}{.\quad}
\usetikzlibrary{calc,patterns,angles,quotes,decorations.pathmorphing}

\newtheorem{lemma}{\bf Lemma}
\newtheorem{theorem}{\bf Theorem}
\newtheorem{proposition}{\bf Proposition}
\newtheorem{remark}{\bf Remark}

\newcommand{\Ri}[1]{\mathbb{R}^{#1}}
\newcommand{\Cont}[1]{\mathcal{C}^{#1}}
\newcommand{\I}[1]{{{\bf I}}_{#1}}
\newcommand{\0}[1]{{{\bf 0}}_{#1}}

\DeclareMathOperator*{\argmin}{arg\,min}

\newcommand{\inv}{^{-1}}
\newcommand{\pinv}{^{\dagger}}
\newcommand{\transp}{^{\mathsf{T}}}
\newcommand{\rank}[1]{{\text{rank}~{ #1}}}

\newcommand{\VHC}{{\mathbf{\Phi}}}
\newcommand{\gc}{{\mathbf{q}}}
\newcommand{\tvc}{{\mathbf{x}_{\perp}}}

\newcommand{\ytvc}{{\mathbf{y}_\perp}}
\newcommand{\ztvc}{{\mathbf{z}_\perp}}
\newcommand{\Dtvc}{{\dot{\mathbf{x}}_{\perp}}}
\newcommand{\Ffunc}{{\mathbf{\Lambda}}}
\newcommand{\nvel}{{\rho}}
\newcommand{\Ds}{\dot{s}}
\newcommand{\DDs}{\ddot{s}}
\newcommand{\sspace}{\mathcal{S}}
\newcommand{\xs}{\mathbf{x}_s}
\newcommand{\pperp}{\mathbf{p}_{\perp}}

\newcommand{\Dp}{\mathbf{D}P}
\newcommand{\DDp}{\mathbf{D}^2P}
\newcommand{\Dps}{\mathbf{D}P_s}
\newcommand{\DDps}{\mathbf{D}^2P_s}

\newcommand{\Om}{\mathbf{\Omega}}
\newcommand{\Oms}{\mathbf{\Omega}_s}
\newcommand{\Pis}{\mathbf{\Pi}_s}

\newcommand{\nomOrb}{\mathbf{\eta}_*}

\newcommand{\hbf}{\mathbf{h}}
\newcommand{\fbf}{\mathbf{f}}
\newcommand{\gbf}{\mathbf{g}}
\newcommand{\Abf}{\mathbf{A}}
\newcommand{\Ubf}{\mathbf{U}}
\newcommand{\Kbf}{\mathbf{K}}
\newcommand{\wbf}{\mathbf{w}}

\newcommand{\Dgc}{{\dot{\mathbf{q}}}}
\newcommand{\DDgc}{{\ddot{\mathbf{q}}}}
\newcommand{\state}{{\mathbf{x}}}
\newcommand{\Dstate}{{\dot{\mathbf{x}}}}
\newcommand{\ac}{{\mathbf{u}}}
\newcommand{\acv}{{\mathbf{v}}}
\newcommand{\Mmat}{\mathbf{M}}
\newcommand{\Cmat}{\mathbf{C}}
\newcommand{\Gmat}{\mathbf{G}}
\newcommand{\Bmat}{\mathbf{B}}

\newcommand{\Fmat}{\mathbf{F}}

\def\withproofs{1}

\title{\LARGE \bf
Excessive Transverse Coordinates   for Orbital Stabilization 
of (Underactuated) Mechanical Systems 
}

\author{Christian Fredrik Sætre$^{1}$, Anton Shiriaev$^{1}$, Stepan Pchelkin$^{1}$,  Ahmed Chemori$^{2}$ 
\thanks{$^{1}$Department of Engineering Cybernetics, NTNU,
        Trondheim, Norway.       {\tt\footnotesize	 \{christian.f.satre,}{\tt\footnotesize	 anton.shiriaev,}{\tt\footnotesize	 stepan.pchelkin }\}{\tt\footnotesize	@ntnu.no}}
\thanks{$^{2}$LIRMM, University of Montpellier, CNRS, Montpellier, France.   {\tt\footnotesize	chemori@lirmm.fr}}
}
\begin{document}

\maketitle

\thispagestyle{plain}
\pagestyle{plain}

\begin{abstract}
Transverse linearization-based approaches have become among the most prominent methods for orbitally stabilizing feedback design in regards to (periodic) motions of underactuated mechanical systems. Yet, in an $n$-dimensional state-space, this requires knowledge of a set of $(n-1)$ independent transverse coordinates, which can be nontrivial to find and whose definitions might vary for different motions (trajectories). In this paper, we consider instead a generic set of \emph{excessive} transverse coordinates which are defined in terms of a particular parameterization of the motion and a projection operator recovering the ``position" along the orbit. We  present a constructive procedure for obtaining the corresponding transverse linearization, as well as state a sufficient condition for the existence of a  feedback controller rendering the desired trajectory (locally) asymptotically orbitally stable. The presented approach is applied to stabilizing oscillations of the underactuated cart-pendulum system about its unstable upright position, in which a novel motion planning approach  based on virtual constraints is utilized for trajectory generation.
\end{abstract}

\begin{keywords}
Underactuated mechanical systems,  orbital stabilization, transverse coordinates, transverse linearization.
\vspace{-0mm}
\end{keywords}
\section{Introduction}
We consider the task of generating feedback controllers that orbitally stabilize  periodic trajectories of  underactuated Euler-Lagrange systems, defined by
\begin{equation}
   \Mmat(\gc)\DDgc+\Cmat(\gc,\Dgc)\Dgc+\Fmat(\gc)\Dgc+\Gmat(\gc)=\Bmat \ac.
    \label{eq:ELeqs}
\end{equation}
 Here $\gc\in \Ri{n_q}$  and $\Dgc=\frac{d}{dt}\gc$  are the generalized coordinates and velocities;   $\ac \in \Ri{n_u}$ is a vector of $n_u (<n_q)$  control inputs;  $\Mmat:\Ri{n_q}\to \Ri{n_q\times n_q}$ is the symmetric, positive definite  inertia matrix; $\Cmat:\Ri{n_q}\times\Ri{n_q}\to \Ri{n_q\times n_q}$ is the matrix of Coriolis and centrifugal terms  that satisfies $\Cmat(\gc,X)Y=\Cmat(\gc,Y)X$ for any $X,Y\in\Ri{n_q}$; $\Fmat:\Ri{n_q}\to \Ri{n_q\times n_q}$ is a  continuously differentiable matrix function of damping and friction terms; $\Gmat:\Ri{n_q}\to \Ri{n_q}$ is the gradient of the system's potential energy; while  $\Bmat\in\Ri{n_q\times n_u}$ is a constant matrix of full rank, i.e. $\rank{\Bmat} =\min(n_q,n_u)$. 

The general problem of stabilizing a predetermined  motion (trajectory) of 
such systems  can  be highly challenging due to both the nonlinear dynamics and the underactuation. For instance, this
prohibits the use of simplifying strategies such feedback linearization, while alternative techniques such as partial feedback linearization \cite{spong1994partial} will result in some remaining internal dynamics, which can be (made) unstable (non-minimum phase), and consequently must  be considered in any control design.

Linearization of the dynamics along a nontrivial solution is also of limited use for the purpose of control design. Indeed,  
unlike an equilibrium point
whose (in-)stability can be  determined simply from the
stability  of the corresponding linearized (first approximation) system,  
the linear variational system  of an autonomous (closed-loop) system evaluated along a periodic solution can never be asymptotically stable, even though the periodic \emph{orbit} is \cite{demir2000floquet}. This well-known fact can for instance be derived from the Andronov-Vitt theorem, which states  that the linear periodic system  corresponding to the linearization along a periodic trajectory   always will have  a non-vanishing, zero-characteristic exponent solution \cite{leonov2006generalization}.

For such solutions, it can therefore  be beneficial to instead consider  the stability of the corresponding orbit (the set of all states along the solution). This is the notion of  \emph{orbital (Poincaré) stability} \cite{leonov2006generalization,urabe1967nonlinear}, in which asymptotic  orbital stability simply means asymptotic convergence to the orbit itself, and not to a specific point in time (moving) along it.

It is well known that a periodic orbit is exponentially stable in the orbital sense if, and only if, the linearization of the dynamics \emph{transverse} to the orbit is exponentially stable \cite{hauser1994converse}. Indeed, this is true for any orbit if the corresponding linearized system is regular \cite{leonov2006generalization,leonov2007time}.  Thus, if one can find $(2n_q-1)$ independent \emph{transverse coordinates}, which vanish on the nominal orbit, and then exponentially stabilize the origin of the corresponding \emph{linearized transverse dynamics} (the first approximation system), then one simultaneously asymptotically stabilizes the trajectory in the orbital sense. 

Finding a set of $(2n_q-1)$ independent transverse coordinates for a particular motion can  be nontrivial though. That is, finding a set of   coordinates $\tvc \in\Ri{2n_q-1}$ together with a scalar variable, $s$, parameterizing the trajectory, such that there exists a (local) diffeomorphism $(\gc,\Dgc) \mapsto (s,\tvc)$.  However, as we will see, there exists a generic choice of $2n_q$ \emph{excessive} transverse coordinates. Since these excessive coordinates, by definition, are dependent on a minimal set of coordinates, the stability of their origin implies the stability of the origin of the minimal coordinates, and consequently the orbital stability of the desired trajectory.

The same set of excessive coordinates we consider in this paper, together with the linearization of their dynamics, has previously been  considered in  \cite{pchelkin2016orbital}  for stabilizing periodic motions of a fully-actuated robot manipulator. Moreover, they were utilized in \cite{saetre2019trajectory}  for the    stabilization of a hybrid walking cycle of a three-link biped robot with two degrees of underactuation, where a particular choice of the parameterizing variable allowed for one coordinate to be trivially  omitted in order to obtain a minimal set of transverse coordinates.

In this paper, we build upon and extend the aforementioned works by presenting several original contributions which provide new insights into excessive transverse coordinates and the linearization of their dynamics. 

The main contributions of the present paper are:
\begin{itemize}
    \item Providing analytical expressions for the transverse linearization  of the excessive transverse coordinates without the need to numerically solve a matrix equation as was required in  \cite{pchelkin2016orbital};
        \item Allowing for the projection operator, which recovers the parameterizing variable of the nominal orbit, to be implicitly defined, and which can depend on all of the system's states, not only its configuration as in \cite{pchelkin2016orbital};
          \item Illustrating that   transverse coordinates need only to  be locally transverse to the flow of the nominal orbit rather than  restricted to being locally orthogonal as  in \cite{hauser1994converse,mohammadi2018dynamic};
    \item Providing an explicit procedure for obtaining an asymptotically orbitally stabilizing feedback controller. 
\end{itemize}

It is also worth noting that the proposed method is  not sensitive to singularities in the  reduced dynamics (see e.g. \cite{shiriaev2008can}). Thus  it can be utilized for the  stabilization of a richer set of trajectories than  the method in \cite{shiriaev2005constructive,shiriaev2010transverse}, such as trajectories whose generating equations have singularities \cite{surov2018new} (see also Sec.~\ref{sec:CP}) and even certain non-periodic trajectories (assuming the linearization is regular).   The method is also easily applicable to systems of any degree of underactuation, as well fully- and  redundantly actuated systems, and has the added benefit that any change in actuation will require minor changes in the presented procedure.

A brief outline of the paper follows.
We begin by defining a set of excessive transverse coordinates for a given trajectory in Sec.~\ref{sec:ETC} and then derive the linearization of their dynamics in Sec.~\ref{sec:TVL}. We then state the main result of this paper in Sec.~\ref{sec:OrbStab} on the form of Theorem~\ref{theorem:MainResult}, which gives sufficient conditions for attaining an orbitally stabilizing controller.  In Sec.~\ref{sec:ProjectionFromImpFuncTh} we provide conditions that allows for the construction of projection operators. While lastly, in Sec.~\ref{sec:CP}, we  illustrate the proposed procedure by stabilizing upright oscillations of the underactuated cart-pendulum system.

\section{Excessive transverse Coordinates}\label{sec:ETC}
Let $\state=[\gc\transp,\Dgc\transp]\transp\in\Ri{2n_q}$ denote the state vector of \eqref{eq:ELeqs}, and suppose a non-trivial (non-vanishing), $T$-periodic trajectory
\begin{equation*}
  \state_*(t,\state_0)=\state_*(t+T,\state_0), \ \state_*(0,\state_0)=\state_0, \ T>0,  
\end{equation*}
is known, as well as the corresponding nominal control input $\ac_*(t,\state_0)$. Further suppose that the corresponding orbit (the set of all states along the trajectory), denoted  $\eta_*$, admits a reparametrization in terms of a (strictly) monotonically increasing scalar variable $s\in \sspace$. We will refer to the parameterizing variable,  $s$, as the the \emph{motion generator} (MG) of the reparameterized trajectory,  defined by
\begin{equation}
\xs(s)=\begin{bmatrix}
    \gc_*(s) \\ \Dgc_*(s)
    \end{bmatrix}
    =\begin{bmatrix}
   \VHC(s) \\ \VHC'(s)\nvel(s)
    \end{bmatrix}
     .
    \label{eq:MGnominalTrajectory}
\end{equation}
 Here $\VHC'(s)=\frac{d}{ds}\VHC(s)$ with  $\VHC:\sspace\to \Ri{n_q}$ 
 at least thrice continuously differentiable, while $\nvel:\sspace\to\Ri{}$ is a $\Cont{2}$-function that recovers the  nominal velocity of the MG along the orbit, i.e. $\Ds_*(t)=\nvel(s(t))>0$, and whose existence  is guaranteed  by the monotonicity of $s$ and the existence of the orbit $\eta_*$,  defined by $\eta_*:=\{\state\in\Ri{2n_q}: \ \state=\xs(s), \ s\in\sspace\}$.  
 
 Note that we will use the  subscript-notation ``$s$"  throughout this paper to denote that a function is evaluated along the trajectory parameterized by the MG, e.g. $h_s(s):=h(\xs(s))$ for any $h:\Ri{2n_q}\to H$ and an arbitrary space $H$. Moreover, we use  $h_s'(s)=\frac{dh_s}{ds}(s)$ for any smooth function $h_s:\sspace \to H$. 
 
Suppose that within some tubular neighbourhood $\mathcal{X}\subset \Ri{2n_q}$ of the orbit $\nomOrb$, the MG, $s$, can be found from a projection of the system states upon the orbit by an operator
\begin{equation}\label{eq:projectionP}
   P:\Ri{2n_q}\to\mathcal{S},\quad \state\mapsto P(\state),  \quad \forall \state \in \mathcal{X},
\end{equation}
which is twice continuously differentiable
in $\mathcal{X}$ and satisfies  $ s=P(\xs(s))$ for all $s\in\mathcal{S}$. We will denote by  $\Dp(\cdot)=\Big[\frac{\partial P}{\partial \gc}(\cdot),\frac{\partial P}{\partial \Dgc}(\cdot)\Big]$ the   Jacobian  of the mapping $P(\cdot)$  and by $\DDp(\cdot)=\frac{\partial^2 P}{\partial \state^2}(\cdot)$  its $2n_q\times2n_q$ symmetric Hessian matrix.

The  idea behind this projection operator is simply that it allows one to project the current state, at least within some tubular neighbourhood, down upon the nominal orbit to recover the ``position"  along it. This then allows one to define some  measure of the distance to this  orbit, which, unlike regular reference tracking, will only depend  on the current state of the system and not on some time-varying reference,  thus giving rise to a completely state-dependent feedback and consequently an autonomous closed-loop system.
More specifically, consider  
\begin{equation}\label{eq:ETVC}
\tvc:=\state-\xs(s),
\end{equation}
which for a particular mapping $P(\cdot)$ are well defined  for all $\state\in\mathcal{X}$. Differentiating \eqref{eq:ETVC} with respect to time leads to
\begin{equation}\label{eq:pureTVD}
    \Dtvc=\left(\I{2n_q}-\xs'(s)\Dp(\state)\right)\Dstate=:\Om(\state)\Dstate,  
\end{equation}
where $\I{m}$ is the $m\times m$ identity matrix.
It  follows that sufficiently close to the orbit, a small variation in the states, $\delta \state$,  relates to a small variation of the coordinates \eqref{eq:ETVC} through
\begin{equation}\label{eq:VariationOmDef}
    \delta \tvc =\Oms(s)\delta \state. 
\end{equation}
Here the (Jacobian) matrix function  $\Oms(s)$ is  of particular interest. We  therefore state some of its key properties next.
\begin{lemma}\label{lemma:Omegas}
    Let $P(\cdot)$ be defined as in \eqref{eq:projectionP} and the curve $x_s:\sspace\to \eta_*$  by \eqref{eq:MGnominalTrajectory}.
    Then for all $s\in \sspace$, the matrix function
    \begin{equation}
    \Oms(s):=\I{2n_q}-\xs'(s)\Dps(s)
    \end{equation}
     is a projection matrix, that is $\Oms^2(s)=\Oms(s)$, and its rank is always $(2n_q-1)$. Moreover, $\Dps(s)$ and $\xs'(s)$ are its left- and right annihilators, respectively. \QEDopen
\end{lemma}

All the properties of $\Oms(\cdot)$ in the above statement are just  straightforward  consequences of the relation $s=P(\xs(s))$, or equivalently $\nvel(s)=\Dps(s)\xs'(s)\nvel(s)$, and hence
\begin{equation}\label{eq:DxsDpRel}
    \Dps(s)\xs'(s)\equiv 1, \quad \forall s\in\sspace.
\end{equation}
Here $\xs'(s)=[{\VHC'}\transp(s),\Ffunc\transp(s)]\transp$  with $\Ffunc(\cdot)$  defined as
\begin{equation}\label{eq:Ffunc}
\Ffunc(s):=\VHC'(s)\nvel'(s)+\VHC''(s)\nvel(s). 
\end{equation}

Note, however, that \eqref{eq:DxsDpRel} does {not} necessarily imply that $\Dps(s)={\xs'}\transp(s)/\|\xs'(s)\|^2_2$ in general. Rather,  let $\theta(s)\in(-\frac{\pi}{2},\frac{\pi}{2})$ denote the angle between $\Dps(s)$ and ${\xs'}\transp(s)$ in their common plane. Then there exists a $\Cont{1}$ vector function $\mathbf{n}_\perp:\sspace \to \Ri{2n_q}$ of unit length within the span of the kernel  of ${\xs'}\transp(s)$, such that by  \eqref{eq:DxsDpRel} and the definition of the  inner product, we have 
\begin{equation}\label{eq:GenDpsDef}
    \Dps(s)=\frac{\xs'(s)\transp}{\|\xs'(s)\|^2_2}+\tan(\theta(s))\frac{\mathbf{n}_\perp(s)\transp}{\|\xs'(s)\|_2}.
\end{equation}
This simple observation is important,  as by Lemma~\ref{lemma:Omegas} and \eqref{eq:VariationOmDef} we  can infer that  $\Oms(s)\delta \tvc=\delta \tvc$,  and hence
\begin{equation*}
\Dps(s)\delta \tvc=\Dps(s)\Oms(s)\delta \tvc\equiv 0    
\end{equation*}
must hold for all $s\in\sspace$. 
This, together with \eqref{eq:GenDpsDef}, thus allows us to conclude that, sufficiently close to the orbit, the coordinates $\tvc$ are orthogonal to $\Dps(s)$ and thus transverse to the flow of the orbit; however, they are not necessarily strictly orthogonal to it. Consequently, they constitute a valid set of transverse coordinates, but  as the matrix function $\Oms(s)$ is not invertible (its rank is always $(2n_q-1)$),   they are an \emph{excessive} set of transverse coordinates. Nevertheless, as is implied by the following statement  (see also \cite[Theorem 3]{pchelkin2016orbital}), if the origin of these coordinates is asymptotically stable, then  so is also the nominal periodic orbit.
\begin{lemma}\label{lemma:minTvcETVCrel}
    Let $\ytvc:\sspace\times \Ri{2n_q}\to\Ri{2n_q-1}$ be a valid  minimal set of transverse coordinates together with a projection operator $P(\cdot)$ as defined in \eqref{eq:projectionP}. That is, $\state\mapsto (s,\ytvc)$ is a local diffeomorphism and  $\ytvc$ vanishes on $\eta_*$. Then the origin of $\ytvc$ is asymptotically stable  if, and only if, the origin of the excessive coordinates $\tvc$ is asymptotically stable.  \QEDopen\if\withproofs0\footnote{The proofs of all the statements in this text are given in the extended version of this paper which is available on the arXiv:  arXiv:1910.00537. }\fi
\end{lemma}

\if\withproofs1
 The proof of Lemma~\ref{lemma:minTvcETVCrel} is stated in Appendix A.
\fi

The value of these excessive transverse coordinates should therefore be evident: given a known solution to \eqref{eq:ELeqs},  they are a valid set of transverse coordinates for \emph{any} parameterization of the form \eqref{eq:MGnominalTrajectory} and \emph{any} projection operator \eqref{eq:projectionP}. They also allow  one to easily change between different sets of coordinates by simply changing either (or both) the parameterization or the projection operator. 

Thus, with the aim of asymptotically stabilizing  the origin of these coordinates, and consequently stabilizing the  orbit, we will show next how one can derive the linearization (first approximation) of their dynamics along the target motion.

\section{Deriving the  transverse linearization}\label{sec:TVL}
Let $\Bmat^\dagger\in\Ri{n_u\times n_q}$ denote a left-inverse of $\Bmat$, that is $\Bmat^\dagger \Bmat=\I{n_u}$, and define the following matrix function:
\begin{equation*}
    \Ubf(\gc,\Dgc,s):=\Mmat(\gc)\Ffunc(s)\nvel(s) +\Cmat(\gc,\Dgc)\Dgc+\Fmat(\gc)\Dgc+\Gmat(\gc).
\end{equation*}
It is not difficult to see that $\Bmat^\dagger\Ubf(\gc,\VHC'(s)\nvel(s),s)$   corresponds to the nominal control input $\ac_*$ when on the nominal orbit. Thus, consider the  feedback transformation
\begin{equation}\label{eq:ContInput}
    \ac=\Bmat^\dagger \hat{\Ubf}(\state,s)+\acv
\end{equation}
where $\hat{\Ubf}:\Ri{2n_q}\times \sspace\to\Ri{n_q}$ is a smooth function satisfying $\hat{\Ubf}(\xs(s),s)\equiv \Ubf(\VHC(s),\VHC'(s)\nvel(s),s)$ for all $s\in\sspace$, while 
 $\acv\in\Ri{n_u}$ is some stabilizing control input to be defined.\footnote{Some natural choices for the function $\hat{\Ubf}$ are $\Ubf(\VHC(s),\VHC'(s)\nvel(s),s)$, $\Ubf(\gc,\VHC'(s)\nvel(s),s)$ or simply $\Ubf(\gc,\Dgc,s)$. }
\begin{proposition}[Excessive transverse linearization]\label{Prop:LTVD}
    With the control law  \eqref{eq:ContInput}, the first approximation (linearization) of the dynamics of the transverse coordinates \eqref{eq:ETVC} along the trajectory \eqref{eq:MGnominalTrajectory} can be written as the constrained (differential-algebraic) linear periodic  system
    \begin{align}\label{eq:LTVD}
         \frac{d}{dt}\delta \tvc &=\Abf_\perp(s)\delta \tvc+\Bmat_\perp(s)\acv 
         \\ \nonumber
         0&=\Dps(s)\delta \tvc
    \end{align}
    where 
    \begin{align*}
    &\hspace{-5mm} \mathbf{A}_\perp(s):= \Oms(s)\mathbf{A}(s)
   -\xs'(s){\xs'}\transp(s)\DDps(s)\nvel(s),
       \\
    &\hspace{-5mm} \mathbf{A}(s) \left.:=\begin{bmatrix}
        \0{n_q} & \I{n_q}
        \\
       {\Mmat(\gc)}\inv\frac{\partial \Tilde{\Ubf}}{\partial \gc}(\state,s)&{\Mmat(\gc)}\inv\frac{\partial \Tilde{\Ubf}}{\partial \Dgc}(\state,s)
    \end{bmatrix}\right\lvert_{\state=\xs(s)} \hspace{-10mm},
    \\
    &\hspace{-4mm} \mathbf{B}_\perp(s):=\left. \Oms(s)\begin{bmatrix} \0{n_q\times n_u} \\ \Mmat(\gc)\inv\Bmat    \end{bmatrix}\right\lvert_{\state=\xs(s)} \hspace{-10mm},
\end{align*}
and $\Tilde{\Ubf}(\state,s):=\Bmat\Bmat\pinv\hat{\Ubf}(\state,s)-\Ubf(\gc,\Dgc,s)$.
\QEDopen
\end{proposition}

\if\withproofs1
The proof of Proposition~\ref{Prop:LTVD} is given in Appendix B. 
\fi
\begin{remark}
The matrix function $\Abf_\perp(s)$ in \eqref{eq:LTVD} is not unique. That is to say, as $\delta \tvc=\Oms(s)\delta \tvc$ and $\Dps(s)\delta\tvc\equiv 0$, the matrix function $\Abf_\perp(s)\Oms(s)+\mathbf{X}(s)\Dps(s)$ would also be a valid choice for any smooth, bounded, vector function $\mathbf{X}:\sspace\to\Ri{2n_q}$.  \QEDopen
\end{remark}
\begin{remark}
For computing  $\Abf(\cdot)$, it can  be useful to note that, by defining  $\ztvc:=\Dgc-\VHC'(s)\nvel(s)$, one has
\begin{align*}
    \Ubf(\gc,\Dgc,s)=&\Ubf(\gc,\VHC'(s)\nvel(s),s)+\Cmat(\gc,\ztvc)\ztvc
    +2\Cmat(\gc,\VHC'(s)\nvel(s))\ztvc+\Fmat(\gc)\ztvc
\end{align*}
 due to $\Cmat(\gc,X)Y=\Cmat(\gc,Y)X$ for any $X,Y\in\Ri{n_q}$. \QEDopen
\end{remark}

The  importance of  Proposition~\ref{Prop:LTVD} is simply  that, by  utilizing the structure of the mechanical system \eqref{eq:ELeqs}, it provides  explicit expressions for the linearized transverse dynamics of the excessive coordinates \eqref{eq:ETVC}  which are  valid for any trajectory of the form \eqref{eq:MGnominalTrajectory}, any feedforward-like input
$\hat{\Ubf}$, and any projection operator  \eqref{eq:projectionP}. 
The statement is of course also true for fully actuated systems ($n_u \equiv n_q$), in which, as then  $\Bmat \Bmat^\dagger=\I{n_q}$, one  has the option of using the (partial-) feedback linearizing-like controller  $\hat{\Ubf}=\Ubf(\gc,\Dgc,s)$, thus resulting in $\Tilde{\Ubf}\equiv \0{}$.

While it is known \cite{pchelkin2016orbital} that the system \eqref{eq:LTVD} can be successfully stabilized in the fully actuated case by a linear feedback of the form $\acv=\Kbf \delta \tvc$ for some constant $\Kbf\in\Ri{n_q\times 2n_q}$, this will in general not be possible  for underactuated systems.  Instead, one must find a smooth matrix function $\Kbf: \sspace \to\Ri{n_u\times 2n_q}$ which varies along the trajectory. We therefore address the issue of how to find such a feedback next.

\section{Stabilization of  the  transverse dynamics}\label{sec:OrbStab}
Since we  consider periodic orbits, for which it is well known that the (asymptotic) stability of the first approximation implies (asymptotic) stability of the nonlinear system (see e.g. \cite{leonov2006generalization}), the following statement naturally holds. 
\begin{lemma}\label{lemma:OrbStab}
    Suppose that there exists a continuously differentiable matrix function $\Kbf:\sspace \to \Ri{n_u\times 2n_q}$ such  $\acv=\Kbf(s)\delta \tvc$ asymptotically stabilizes the origin of  \eqref{eq:LTVD}. Then the control law \eqref{eq:ContInput} with $\acv=\Kbf(P(\state))\tvc$ renders the nominal orbit $\eta_*$  asymptotically stable, and consequently the desired solution $\state_*(t,\state_0)$ asymptotically orbitally stable. \QEDopen
\end{lemma}

The question then arises as to how one can find such a matrix function $\Kbf(\cdot)$. If, for instance, the pair $\left(\mathbf{A}_\perp(\cdot),\mathbf{B}_\perp(\cdot)\right)$ were  stabilizable, then it is known (see e.g. \cite{yakubovich1986linear}) that an exponentially stabilizing controller would be given by  
\begin{equation}\label{eq:LQRcont}
    \acv=-\mathbf{\Gamma}^{-1}\mathbf{B}_{\perp}\transp(s)\mathbf{R}(s)\tvc,
\end{equation}
where the matrix function $\mathbf{R}(\cdot)$ is the symmetric, positive semi-definite  solution of the differential Riccati equation 
 \begin{align*}
    &\dot{\mathbf{R}}(s)+\mathbf{A}_{\perp}\transp(s)\mathbf{R}(s)+\mathbf{R}(s)\mathbf{A}_{\perp}(s)+\mathbf{Q}
    +\kappa \mathbf{R}(s)-\mathbf{R}(s)\mathbf{B}_{\perp}(s)\mathbf{\Gamma}^{-1}\mathbf{B}_{\perp}\transp(s)\mathbf{R}(s)=0,
\end{align*}
for some $\mathbf{\Gamma}^{}=\mathbf{\Gamma}\transp\succ 0$, $\mathbf{Q}=\mathbf{Q}\transp\succ 0$ and $\kappa\ge0$.

Unfortunately, however, it can be shown \cite[Proposition 9]{saetre2019excessive} (see also Sec. 4.2 in \cite{demir2000floquet}) that the pair $\left(\mathbf{A}_\perp(\cdot),\mathbf{B}_\perp(\cdot)\right)$ can never be stabilizable even though the origin of the system \eqref{eq:LTVD} can be asymptotically (exponentially) stabilized. More precisely, the system $\dot{\wbf}=\Abf_\perp(s)\wbf+\Bmat_\perp(s)\acv$,  corresponding to \eqref{eq:LTVD} without the transversality condition $\Dps(s)\wbf=0$, always has a non-vanishing solution in the direction of $\xs'(s)$, regardless of the control input $\acv$. 

Although this implies that no solution $\mathbf{R}(\cdot)$ to the above (periodic) differential Riccati equation can exist, we can instead try to find a solution of a modified Riccati equation, which, if found,  allows for the generation of a stabilizing controller. 
This leads us  to the main result of this paper. 
\begin{theorem}\label{theorem:MainResult}
Suppose there exists a  symmetric, positive semi-definite solution $\mathbf{R}_\perp(\cdot)$ to the following modified periodic differential Riccati equation:
 \begin{align} \label{eq:RDE}
    &\Oms\transp(s)\Big[{\mathbf{R}}_\perp'(s)\nvel(s)+\mathbf{A}_{\perp}\transp(s)\mathbf{R}_\perp(s)+\mathbf{R}_\perp(s)\mathbf{A}_{\perp}(s)+\mathbf{Q}  
    +\kappa\mathbf{R}_\perp(s)-\mathbf{R}_\perp(s)\mathbf{B}_{\perp}(s)\mathbf{\Gamma}^{-1}\mathbf{B}_{\perp}\transp(s)\mathbf{R}_\perp(s)\Big]\Oms(s)=0
\end{align}
with ${\mathbf{R}}_\perp'(s)=\frac{d}{ds}{\mathbf{R}}_\perp(s)$ and for some $\mathbf{\Gamma}=\mathbf{\Gamma}\transp\succ 0$, $\mathbf{Q}=\mathbf{Q}\transp\succ 0$ and $\kappa\ge0$.
Then taking
\begin{equation*}
    \acv=-\mathbf{\Gamma}^{-1}\mathbf{B}_{\perp}\transp(s)\mathbf{R}_\perp(s)\tvc \quad  \text{with}\quad s=P(\state)
\end{equation*}
 in  \eqref{eq:ContInput}
renders the periodic orbit  of the mechanical system \eqref{eq:ELeqs} corresponding to \eqref{eq:MGnominalTrajectory}  locally  asymptotically stable.  \QEDopen
\end{theorem}

\if\withproofs1
The proof of Theorem~\ref{theorem:MainResult} is given in Appendix C.
\fi

Naturally, a solution to the projected periodic differential Riccati equation \eqref{eq:RDE} can only exist if the pair $(\Abf_\perp(s),\Bmat_\perp(s))$ is stabilizable on the set of solutions satisfying the condition $\Dps(s)\delta \tvc \equiv 0$. 
However, the question of the existence  of solutions of this equation is, to our best knowledge, unknown. Although should a solution exist, it is likely not to be unique as    the example considered in \cite{saetre2019excessive} demonstrates. Nevertheless, as we will see in the example of Sec.~\ref{sec:CP}, we have been able to find accurate approximate solutions using numerical methods.

\section{On Obtaining a Projection operator}\label{sec:ProjectionFromImpFuncTh}
So far, we have  assumed that a projection operator of the form $s=P(\state)$, with the mapping $P(\cdot)$ as defined in \eqref{eq:projectionP},  is known and given as an explicit equation  which is well defined within some neighbourhood of the orbit. Yet, this might  not necessarily always be the case. That is to say, if one has found a feasible trajectory  of the system \eqref{eq:ELeqs}  parameterized on the form \eqref{eq:MGnominalTrajectory}, the corresponding motion generator might only be known as a function of time, i.e.  $s=s(t)$; indeed,  $s=t$ is of course the most commonly used parameterization of trajectories. 
 Therefore, we will briefly  show  next how one can generate  projection operators  given only  knowledge of the nominal trajectory and the time evolution of its parameterizing variable.
 
 In this regard, the following statement follows directly from the implicit function theorem. 
 \begin{proposition}\label{prop:FindProjOp}
 Assume that on a given subarc of the trajectory \eqref{eq:MGnominalTrajectory}, denoted $\sspace_k\subseteq \sspace$, there exists a smooth function $h_k: \Ri{2n_q}\times \sspace_k \to \Ri{}$  satisfying
\begin{equation}\label{eq:h_func}
	h_k(\state_s(s),s)\equiv 0 \ \ \text{and} \ \ \frac{\partial h_k}{\partial s}(\state_s(s),s)\neq 0, \ \ \forall s\in\sspace_k.
\end{equation}
 Then,  in some nonzero, tubular neighbourhood $\mathcal{X}_k\subset \Ri{2n_q}$ of the orbit $\eta_*$, there exists a function $P_k: \mathcal{X}_k \to \sspace_k$ such that for all $\state\in\mathcal{X}_k$,  we have $h_k(\state,P_k(\state))=0$ as well as 
 \begin{equation*}
\Dp_k(\state)=-{\left(\frac{\partial h_k}{\partial s}(\state,P_k(\state))\right)}^{-1}\frac{\partial h_k}{\partial \state}(\state,P_k(\state)). \quad   \QEDopen
 \end{equation*}

 \end{proposition}
 
It follows that if a function $h_k(\cdot)$ satisfying the conditions of Proposition~\ref{prop:FindProjOp}  for   $\state(t)\in\mathcal{X}_k$ is found, then one can take 
 \begin{equation*}
	s=P_k(\state(t))
 \end{equation*}
 as the projection of the states at time $t$ onto the $\sspace_k$ subarc of the orbit \eqref{eq:MGnominalTrajectory}.  
Such a function $h(\cdot)$ can often be found satisfying \eqref{eq:h_func} on the whole trajectory. For example, the solution to the following implicit equation  
\begin{equation*}
  s=\argmin_{s\in\mathcal{S}} \left(\state(t)-\state_s(s)\right)\transp \mathbf{V}(s)\left(\state(t)-\state_s(s)\right)
\end{equation*}
with $\mathbf{V}:\sspace \to \Ri{2n_q\times 2n_q}$ a continuous, symmetric, positive semi-definite matrix functions  satisfying ${\xs'}\transp(s)\mathbf{V}(s)\xs'(s)>0$ for all $s\in\sspace$, is a generic choice for any trajectory of the form \eqref{eq:MGnominalTrajectory}.
Indeed, with  $\mathbf{V}=\I{2n_q}$, this choice, which  has been  considered several times before  in relation to  stability analysis of autonomous dynamical systems (see e.g. \cite{leonov2006generalization,hartman1962global,borg1960condition,hauser1994converse}), results in  the condition $\xs'(s)\transp \tvc\equiv 0$, and hence
\begin{equation*}
    \Dp(\state)=\frac{{\xs'}\transp(s)}{\|\xs'(s)\|^2_2-\xs''(s)\tvc}.
\end{equation*}
Consequently, the projection \eqref{eq:projectionP} can  be found numerically by, for instance,  a few iterations of Newton's method. 
\section{Example: Upright Oscillations of the Cart-Pendulum System}\label{sec:CP}
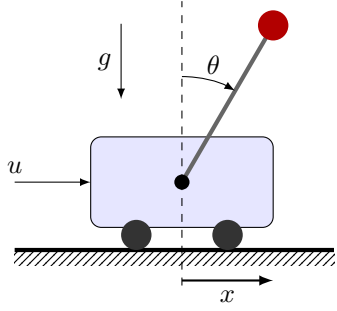
\begin{figure}
    \centering
    \begin{tikzpicture}[scale=2.]
	\coordinate (origo) at (0,0);
	\coordinate (xdir)  at (2.5,0);
	\draw[ultra thick]  ($ (origo) +(0.4,0)  $) -- (xdir);
	\fill[step=2cm,pattern = north east lines] ($ (origo) +(0.4,-0.1)  $) rectangle ($ (origo) + (xdir) $);
	\coordinate (xc) at (1.5,0);
	\coordinate (yc) at (0,0.45);     	
	\coordinate (width) at (0.6,0);  
	\coordinate (heigth) at (0,0.3); 
	\coordinate (S) at ($(xc)+(yc)$);
	\coordinate (pendu) at ($(S)+ (60:1.2cm)$);
	\coordinate (abovec) at ($(xc)+(yc)+(0,1.2)$);
	\draw[fill=blue!10!white,rounded corners] ($(xc)-(width)+(yc)-(heigth)$)rectangle($(xc)+(width)+(yc)+(heigth)$);
    \draw[ultra thick,white!40!black] (S)  -- (pendu);
    \draw[dashed] ($(xc)+(yc)$) -- (abovec) ;
    \draw[dashed] ($(xc)+(yc)$) -- ++(0,-0.7) node[below](test) {};
    \draw[-latex,thick] ($(test)+(0,0.11)$) -- ++(0.6,0)node[midway,below]{$x$};
    \draw pic [" $\theta$",draw, latex-,  angle eccentricity=1.15,angle radius=1.4cm] {angle = pendu--S--abovec};
	\draw[-latex] (1.1,1.5)--(1.1,1.)node[midway,left]{$g$};
	\fill[white!20!black] ($(xc)+([scale=0.5]width)+(0,0.1)$) circle (0.1);
	\fill[white!20!black] ($(xc)-([scale=0.5]width)+(0,0.1)$) circle (0.1);
	\draw[latex-] ($(S)-(width)$)--($(S)-(width)-(0.5,0)$)node[above]{$u$};
	\fill (S) circle (0.05);
	\fill[red!70!black] (pendu) circle (0.1);
\end{tikzpicture}
    \caption{Schematic of the cart-pendulum system, which consists of an unactuated pendulum attached to an actuated cart  driven along the horizontal by an external force $u$.}
    \label{fig:CartPendulumSys}
\end{figure}

We will now illustrate the procedure outlined in Sec.~\ref{sec:TVL}-\ref{sec:OrbStab} by stabilizing oscillations around the unstable upright equilibrium of the cart-pendulum system. To keep  the derivation short, we consider  unit masses, and consider the pendulum bob to be a point mass, while  its rod is considered to be massless and of unit length. With the generalized coordinates $\gc=[x,\theta]\transp$ as defined  according to  Fig.~\ref{fig:CartPendulumSys}, the equations of motion of the system are 
\begin{subequations}\label{eq:CPsys}
\begin{align}\label{eq:CPeq1}
    2\ddot{x}+\cos(\theta)\ddot{\theta}-\sin(\theta)\dot{\theta}^2&=u,\\
    \ddot{\theta}+\cos(\theta)\ddot{x}-g\sin(\theta)&=0,\label{eq:CPeq2}
\end{align}
\end{subequations}
where $g=\SI{9.81}{\meter\per\second\squared} $ is the gravitational acceleration.

This task  has  previously  been considered in \cite{shiriaev2005constructive}  utilizing the  \emph{virtual constraints approach}, where a feasible trajectory of the nonlinear system \eqref{eq:CPsys} was generated in the following way. Under the assumption that along  a nominal trajectory of the system a set of relations of the form $(x_*,\theta_*)\transp=\VHC(s)=[\phi_1(s),\phi_2(s)]\transp$ are kept invariant, one can write \eqref{eq:CPeq2} as 
\begin{equation}\label{eq:cp-alpha-beta-gamma}
    \underbrace{\left(\phi_2'+\phi_1'\cos(\phi_2)  \right)}_{\alpha(s)}\DDs+\underbrace{\left(\phi_2''+\phi_1''\cos(\phi_2)\right)}_{\beta(s)}\Ds^2\underbrace{-g\sin(\phi_2)}_{\gamma(s)}=0.
\end{equation}
This  constrains  the time-evolution of the motion generator, $s$, for the particular choice of $\VHC(s)$ and its initial velocity $\Ds_0=\Ds(0)$. Moreover, the nominal velocity $\nvel(s_*(t))=\Ds_*(t)$ can be found as \eqref{eq:cp-alpha-beta-gamma} is integrable \cite{shiriaev2005constructive}; indeed,  the equality
\begin{align}\label{eq:intfunc}
    \frac{1}{2}&\text{exp}\left\{\int_{s_0}^{s} \frac{2\delta(\sigma)}{\alpha(\sigma)}d\sigma\right\}{\alpha^2(s)}\nvel^2(s)-\frac{1}{2}{\alpha^2 (s_0)}\nvel^2(s_0) 
    +\int_{s_0}^s\text{exp}\left\{\int_{s_0}^{\tau} \frac{2\delta(\sigma)}{\alpha(\sigma)}d\sigma\right\}\alpha(\tau)\gamma(\tau)d\tau\equiv 0
\end{align}
must hold, 
where  $\delta(s):=\beta(s)-\frac{d}{ds}\alpha(s)$.
Consequently, the nominal control input $u_*=u_*(s)$ can be found from \eqref{eq:CPeq1}.

In \cite{shiriaev2005constructive}, the holonomic relations $\VHC(\theta)=[-1.5\sin(\theta),\theta]\transp$ were utilized, i.e. the motion generator is simply $s=\theta$, which results in a center at the equilibrium $\theta=\dot{\theta}=0$. 

\begin{figure}
    \centering
    \includegraphics[width=0.5\linewidth]{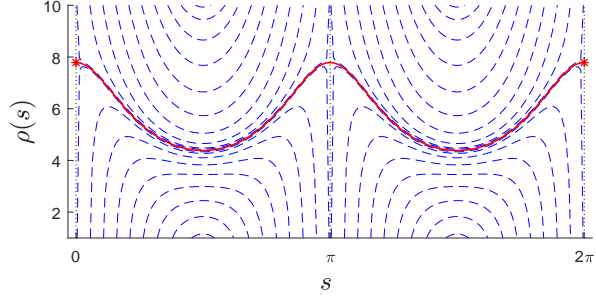}
    \caption{Solutions of  \eqref{eq:intfunc} with the parameterization \eqref{eq:CP_param}. The red curve represents the unique (positive) solution over the interval $[0,2\pi)$ for
    the case of $a_2 = 0.5$.}
    \label{fig:CP_zeta}
\end{figure}

While $s=\theta$ is clearly a convenient choice in this case, it is not consistent with our parameterization \eqref{eq:MGnominalTrajectory} as we require $\Ds_*>0$.  Thus, with  $s\in[0,2\pi)$ and  $\Ds_*=\nvel(s)>0$, let us instead consider 
\begin{equation}\label{eq:CP_param}
\VHC(s)=\big[-1.5\sin\big(a_2\cos(s)\big),a_2\cos(s)\big]\transp,
\end{equation}
which, as we will see, is not holonomic as the projection $P(\cdot)$ will depend on (some of) the generalized velocities. 
 Hence, we can pick $a_2$ to get the appropriate amplitude of the oscillations of the pendulum, compared to \cite{shiriaev2005constructive}  in which the amplitude was determined by the initial conditions $(\theta_0,\dot{\theta}_0)$. Furthermore, while the parameterization in \cite{shiriaev2005constructive} results in a family of periodic solutions around the equilibrium, there exists a unique function $\nvel(s)$ for each choice of $a_2$ in the parameterization \eqref{eq:CP_param}. For example, the unique (positive) solution for the case of $a_2=0.5$ is the  red curve highlighted in Fig.~\ref{fig:CP_zeta}.

It is here worth to note that, even though $\alpha(s_s)\equiv 0$ for $s_s\in\{0,\pi,2\pi\}$, i.e. $s_s$ are singular points\footnote{Note here that the type  of (simple) singularities  presented here are just a product of the choice of parameterization and not due to the non-uniqueness of (phase-space) solutions as  in \cite{surov2018new}.} of the equation \eqref{eq:cp-alpha-beta-gamma}, the solution $\nvel(s)$ of \eqref{eq:intfunc} is well defined  over the interval  $[0,2\pi)$ if we take $\nvel(s_0)$ satisfying $\beta(s_s)\nvel^2(s_s)+\gamma(s_s)=0$.   Therefore, unlike most existing methods (see, e.g. \cite{shiriaev2005constructive}) which require $\alpha(s)\neq 0$ for all $ s\in \sspace$, 
the existence of such singularities is irrelevant for our method.

Now, with the parameterization \eqref{eq:CP_param}, it is clear that we get $\dot{\theta}_*(s)=\phi_2'(s)\nvel(s)=-a_2\sin(s)\nvel(s)$. Hence we can simply find $s$ as the root of the implicit equation
\begin{equation*}
   h(\state,s)=s-\text{arctan2}\left({\dot{\theta}}/{(-\theta\nvel(s))}\right)=0,
\end{equation*}
with $\text{arctan2}(\cdot)$ denoting the four-quadrant arctangent function, thus letting us  utilize the method outlined in Section~\ref{sec:ProjectionFromImpFuncTh}.

\begin{figure}
    \centering
        \begin{subfigure}[b]{\linewidth}
        \centering
        \includegraphics[width=0.6\linewidth]{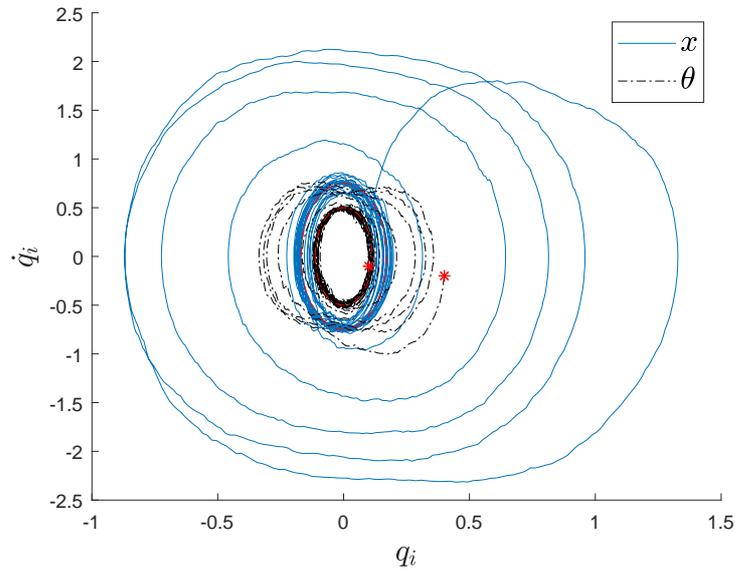}
        \caption{Phase portraits of the \textcolor{blue}{\bf cart (--)}  and the \textcolor{black}{\bf pendulum (-$\cdot$-)}.}
        \end{subfigure}
    \begin{subfigure}[b]{\linewidth}
    \centering
        \includegraphics[width=0.6\linewidth]{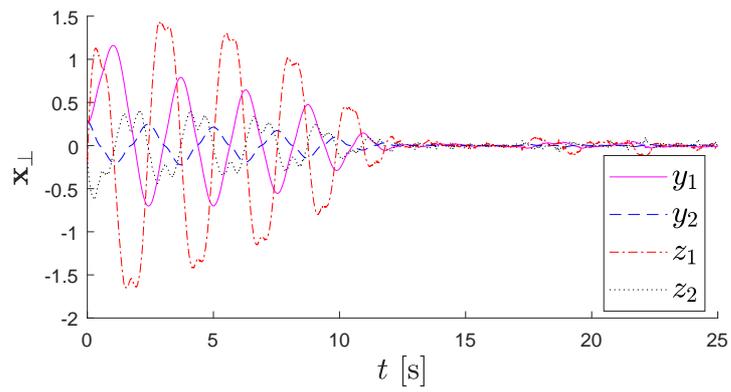}
        \caption{Evolution of the transverse coordinates versus time.}
        \end{subfigure}
        \begin{subfigure}[b]{\linewidth}
        \centering
    \includegraphics[width=0.6\linewidth]{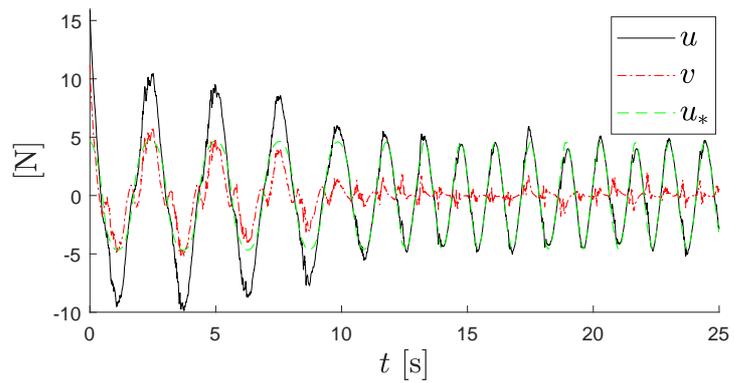}
    \caption{Evolution of the control input signals versus time.}
    \end{subfigure}
    \caption{Results from simulating the cart-pendulum system with  perturbed initial conditions and noisy measurements.}
    \label{fig:CP_sim}
\end{figure}

In order to demonstrate the proposed control scheme, we found that taking $a_2=0.1129$ resulted in a periodic orbit very close to the one considered in \cite{shiriaev2005constructive}.
 Figure~\ref{fig:CP_sim}  shows the results  from  simulating the system  with  the initial conditions ($x_0,\theta_0,\dot{x}_0,\dot{\theta}_0)=(0.1,0.4,-0.1,-0.2)$
and with white noise added to the measurements. The system is seen to  converge  to the orbit after approximately $\SI{13}{\second}$.
Here  a feedback LQR-controller of the form \eqref{eq:LQRcont}  was used, which was generated by solving \eqref{eq:RDE} with $\mathbf{Q}=\I{4}$, $\Gamma=\kappa=0.1$. This was achieved using the semi-definite programming method of \cite{gusev2016sdp} with a trigonometric polynomial of order 40 and utilizing the YALMIP toolbox \cite{lofberg2004yalmip} and the SDPT3 solver \cite{tutuncu2003solving}. The resulting solution satisfied \eqref{eq:RDE} within a maximum error norm of less than $2\times 10^{-4}$ for all $s\in[0,2\pi)$.

\section{Concluding remarks and future work}
In this paper, we have introduced a generic set of excessive transverse coordinates for the purpose of asymptotically orbitally stabilizing periodic trajectories of underactuated Euler-Lagrange systems. 
  We have provided analytical expressions for the corresponding  transverse linearization of these coordinates, which are valid regardless of the choice of parameterization of the trajectory, of the choice of the feedforward-like control input, as well as regardless of the choice of projection operator. In addition, we have  derived a sufficient condition  for the existence of a stabilizing controller for this constrained linear system,   allowing for the construction of a feedback controller rendering the desired periodic motion asymptotically (exponentially) stable in the orbital sense. 
The proposed  scheme was applied to the task of stabilizing oscillations of the cart-pendulum system  around its unstable upright position and was successfully tested in simulation. This example also illustrated that the proposed methodology can, unlike most other methods,  be used to stabilize trajectories for which the reduced dynamics  have singular points.  Experimental validation  of the proposed scheme is currently being pursued.

\section*{APPENDIX}
\section*{Appendix A. proof of Lemma~\ref{lemma:minTvcETVCrel}}
It here suffices to show that the asymptotic stability of the variation $\delta \tvc$ is  equivalent to that of $\delta \ytvc$. Tho this end, we note that by the hypothesis  that the mapping $(s,\ytvc)\mapsto \state$ is a diffeomorphism in a neighbourhood of $\eta_*$, it follows that the Jacobian of $\ytvc$ evaluated along $\xs(s)$, that is
\begin{equation*}
\mathbf{D} \ytvc(s,\xs(s))=\frac{\partial \ytvc}{\partial \state}(s,\xs(s))+\frac{\partial \ytvc}{\partial s}(s,\xs(s))\Dps(s),
\end{equation*}
has full (row) rank for all $s\in \sspace$. But as $\ytvc(s,\xs(s))\equiv 0$, we have $\frac{d \ytvc}{d\state}(s,\xs(s))\xs'(s)\equiv \0{}$, such that by \eqref{eq:DxsDpRel} and by defining $\Pis(s):=\frac{\partial \ytvc}{\partial \state}(s,\xs(s))$, we obtain the relation
\begin{equation*}
    \Pis(s)\xs'(s)=-\frac{\partial \ytvc}{\partial s}(s,\xs(s)).
\end{equation*}
This further implies that
\begin{equation*}
    \delta \ytvc =  \mathbf{D} \ytvc(s,\xs(s))\delta \state=\Pis(s)\Oms(s)\delta \state,
\end{equation*}
such  that from \eqref{eq:VariationOmDef}, i.e. $\delta \tvc=\Oms(s)\delta \state$, we obtain
\begin{equation}\label{eq:deltaMtvcDeltaTvc}
    \delta \ytvc=\Pis(s)\delta \tvc.
\end{equation}

Therefore, from Lemma~\ref{lemma:Omegas} and the fact that $\Pis(s)\Oms(s)$ is of full rank ($2n_q-1$), it follows from \eqref{eq:deltaMtvcDeltaTvc} that $\|\delta \ytvc\|\to 0$ as $t\to \infty$ only if $\|\delta \tvc\|\to 0$.
Thus it just remains to show that the converse is true as well,  namely that $\|\delta \tvc\|\to 0$ as $t\to \infty$ only if $\|\delta \ytvc\|\to 0$.

Towards this end, take  $\pperp:\sspace\to\Ri{2n_q\times (2n_q-1)}$ to be some differentiable  basis of the the kernel of $\Dps(s)$. As then $\pperp(s)=\Oms(s)\pperp(s)$ and  $\rank[\Pis(s)\Oms(s)]=2n_q-1$, it follows that  $\Pis(s)\pperp(s)$ is invertible for all $s\in\sspace$. Hence
\begin{equation*}
    \begin{bmatrix} \Dps(s) \\ \Pis(s)\Oms(s) \end{bmatrix}\begin{bmatrix} \xs'(s) & \pperp(s)\left(\Pis(s)\pperp(s)\right)\inv \end{bmatrix}=\I{2n_q-1},
\end{equation*}
which implies that
\begin{equation*}
    \delta \state =  \pperp(s)\left(\Pis(s)\pperp(s)\right)\inv\delta \ytvc+\xs'(s) \delta s.
\end{equation*}
Left-multiplying the above equation by $\Oms(s)$, we obtain 
\begin{equation*}
    \delta \tvc=\pperp(s)\left(\Pis(s)\pperp(s)\right)\inv\delta \ytvc.
\end{equation*}
It follows that $\|\delta \tvc\|\to 0$ only if  $\|\delta \ytvc\|\to 0$. 
 \QED 
\section*{Appendix B. Proof of Proposition~\ref{Prop:LTVD}}
Let $\ztvc:=\Dgc-\VHC'(s)\nvel(s)$ and  note that  the dynamics of the system \eqref{eq:ELeqs} then can be rewritten on the form
\begin{equation*}
    \Dstate=\xs'(s)\nvel(s)+\begin{bmatrix} \ztvc \\ -\Mmat(\gc)\inv \Ubf(\gc,\Dgc,s)
    \end{bmatrix}+\begin{bmatrix} \0{n_q\times n_u} \\ \Mmat(\gc)\inv\Bmat
    \end{bmatrix}\ac.
\end{equation*}
If the  controller is taken as in \eqref{eq:ContInput}, we obtain 
\begin{equation*}
    \Dstate=\xs'(s)\nvel(s)+\fbf_{\tilde{U}}(\state,s)+\begin{bmatrix} \0{n_q\times n_u} \\ \Mmat(\gc)\inv\Bmat
    \end{bmatrix}\acv,
\end{equation*}
where
\begin{align*}
    &\fbf_{\tilde{U}}(\state,s):=\begin{bmatrix} \0{n_q} \\ \Mmat(\gc)\inv\left[\Bmat\Bmat^\dagger\hat{\Ubf}(\state,s)-\Ubf(\gc,\Dgc,s) \right]
    \end{bmatrix}.
\end{align*}
Note here that  $\Bmat\Bmat^\dagger\hat{\Ubf}(\state_s(s),s)-\Ubf(\VHC(s),\VHC'(s)\nvel(s),s)\equiv 0$.

Consider  now \eqref{eq:pureTVD}; that is $\Dtvc=\Om(\state)\Dstate$.
In order to linearize this system along the orbit, we note that for a differentiable function $\hbf:\Ri{2n_q}\to\Ri{2n_q}$ which,  for all $s\in\sspace$ satisfies $\hbf(\xs(s))\equiv \0{2n_q\times 1}$, then  the relations
\begin{align*}
    \frac{\partial \hbf}{\partial s}(\xs(s))\equiv 0 \quad \text{and}\quad 
    \frac{\partial \hbf}{\partial \tvc}(\xs(s))\Oms(s)=\frac{\partial \hbf}{\partial \state}(\xs(s))
\end{align*}
always hold \cite{pchelkin2016orbital}.
Thus, if we write $\Dstate=\fbf(\state)+\gbf(\state)\acv$, then the matrix $\Bmat_\perp(s)=\Oms(s)\gbf(\xs(s))$ follows from the fact that $\Dtvc$ is affine in the control input $\acv$; whereas the matrix $\Abf_\perp(s)$ must be a solution of the matrix equation 
\begin{equation*}
    \Abf_\perp(s)\Oms(s)=\Oms(s)\frac{\partial \fbf}{\partial \state}(\xs(s))-\Xi(s)\nvel(s)
\end{equation*}
with $\Xi(s):=\xs'(s)\xs'(s)\transp\DDps(s)+\xs''(s)\Dps(s)$. However, as $\Oms(s)\delta \tvc=\delta\tvc$ and $\Dps(s)\delta \tvc\equiv 0$, we must have $\Abf_\perp(s)\delta \tvc=\Abf_\perp(s)\Oms(s)\delta \tvc$; hence, one can simply take
$\Abf_\perp(s)=\Oms(s)\frac{\partial \fbf}{\partial \state}(\xs(s))-\Xi(s)\nvel(s)$. 
Lastly, using that $\frac{\partial h_s(s)}{\partial \state}\delta \tvc=\frac{\partial h_s(s)}{\partial s}\Dps(s)\delta \tvc\equiv 0$ for any $h_s(s):=h(\xs(s))$, we see that $\Abf_\perp(s)$ in \eqref{eq:LTVD} is  a solution to this matrix equation. \QED

\section*{Appendix C. Proof of Theorem~\ref{theorem:MainResult}}
We  will show that if a solution to \eqref{eq:RDE} exists, then the system \eqref{eq:LTVD} is rendered asymptotically stable by controller
\begin{equation*}
    \acv=-\mathbf{\Gamma}^{-1}\mathbf{B}_{\perp}\transp(s)\mathbf{R}_\perp(s)\delta\tvc,
\end{equation*}
letting us utilize Lemma~\ref{lemma:OrbStab}. 
Towards this end, consider the following Lyapunov function candidate:
\begin{equation*}
    V=\delta \state_{\perp}\transp \mathbf{R}_\perp(s)\delta \tvc.
\end{equation*}
Due to the condition $\Dps(s)\delta \tvc \equiv 0$, and consequently $\delta \tvc=\Oms(s)\delta \tvc$, it  can be equivalently rewritten as
\begin{equation*}
    V=\delta \state_{\perp}\transp\Oms\transp(s) \mathbf{R}_\perp(s)\Oms(s)\delta \tvc.
\end{equation*}
Now note that
\begin{equation*}
    \dot{\Om}_s(s)=-[\xs''(s)\Dps(s)+\xs'(s){\xs'(s)}\transp \DDps(s)]\nvel(s).
\end{equation*}
Hence
\begin{align*}
    &\dot{V}=-2\delta \state_{\perp}\transp\Big[\Oms\transp \mathbf{R}_\perp\xs'' \Dps \nvel \Big]\delta \tvc+\delta \state_{\perp}\transp\Big[ \Oms\transp\dot{\mathbf{R}}_\perp\Oms +
    \mathbf{A}_{\perp}\transp\mathbf{R}_\perp\Oms+\Oms\transp\mathbf{R}_\perp\mathbf{A}_{\perp}
    -2\Oms\transp \mathbf{R}_\perp\mathbf{B}_{\perp}\mathbf{\Gamma}^{-1}\mathbf{B}_{\perp}\transp\mathbf{R}_\perp\Oms
    \big]\delta \tvc 
\end{align*}
which, by  the fact that $\Oms^2(s)=\Oms(s)$, is equivalent to
\begin{align*}
    \dot{V}=\delta \state_{\perp}\transp\Oms\transp\Big[ \dot{\mathbf{R}}_\perp
    &+ 
    \mathbf{A}_{\perp}\transp\mathbf{R}_\perp+\mathbf{R}_\perp\mathbf{A}_{\perp} 
    -2 \mathbf{R}_\perp\mathbf{B}_{\perp}\mathbf{\Gamma}^{-1}\mathbf{B}_{\perp}\transp\mathbf{R}_\perp
    \big]\Oms\delta \tvc ,
\end{align*}
where  $\dot{\mathbf{R}}_\perp(s)={\mathbf{R}}_\perp'(s)\nvel(s)$.
Thus by \eqref{eq:RDE}, we obtain
\begin{align*}
    \dot{V}=&\delta \state_{\perp}\transp\Oms\transp\Big[-\mathbf{Q}-\kappa \mathbf{R}_\perp-\mathbf{R}_\perp\mathbf{B}_{\perp}(s)\mathbf{\Gamma}^{-1}\mathbf{B}_{\perp}\transp\mathbf{R}_\perp
    \big]\Oms\delta \tvc\\
    &=\delta \state_{\perp}\transp\Big[-\mathbf{Q}-\kappa \mathbf{R}_\perp-\mathbf{R}_\perp\mathbf{B}_{\perp}(s)\mathbf{\Gamma}^{-1}\mathbf{B}_{\perp}\transp\mathbf{R}_\perp
    \big]\delta \tvc.
\end{align*}
Consequently, for all $\|\delta \tvc\| \neq 0$ satisfying $\Dps(s)\delta \tvc\equiv 0$, we have $\dot{V}<0$, 
which implies  asymptotic stability of the origin of \eqref{eq:LTVD}. \QED

\section*{ACKNOWLEDGMENT}
This work was supported by the Research Council of Norway, project numbers 262363/294538 and MEAE/MESRI French ministries within PHC AURORA Program.


\bibliographystyle{IEEEtran}
\bibliography{IEEEabrv,references.bib}


\end{document}